\documentclass[aps,prl,showpacs,twocolumn,groupedaddress]{revtex4}
\usepackage{amsbsy,amssymb,amsmath,bm}
\usepackage{graphicx}

\input{epsf}

\begin{document}
\title {Collective quantum coherent oscillations in a globally coupled array of qubits}
\author{P. A. Volkov$^{1}$ and M. V. Fistul$^{2,3}$}

\affiliation {$^{1}$ Moscow Institute of Physics and Technology,
Dolgoprudny, 141700 Moskovskii region, Russia\\
$^{2}$ Theoretische Physik III, Ruhr-Universit\"at Bochum, D-44801 Bochum, Germany \\
$^{3}$ Theoretical Physics and Quantum Technologies Department, Moscow Institute for Steel and Alloys,
119049 Moscow, Russia
 }

\date{\today}
\begin{abstract}
We report a theoretical study of coherent collective quantum dynamic effects in an array of $N$ qubits (two-level systems) incorporated into a low-dissipation resonant cavity. Individual qubits are characterized by energy level differences $\Delta_i$ and we take into account a spread of parameters $\Delta_i$. Non-interacting qubits display coherent quantum beatings with $N$ different frequencies, i.e. $\omega_i=\Delta_i/\hbar$ . Virtual emission and absorption of cavity photons provides a long-range interaction between qubits. In the presence of such interaction we analyze quantum correlation functions of individual qubits $C_i(t)$ to obtain \emph{two collective quantum-mechanical coherent oscillations}, characterized by frequencies $\omega_1=\bar{\Delta}/\hbar$ and $\omega_2=\tilde{\omega}_R$, where $\tilde{\omega}_R$ is the resonant frequency of the cavity renormalized by interaction. The amplitude of these oscillations can be strongly enhanced in the resonant case when $\omega_1 \simeq~\omega_2$.
\end{abstract}


\maketitle

Great attention is devoted to theoretical and experimental studies of various superconducting qubits \cite{QM2,Squbits,qubits}. It can be small and large Josephson junctions (charge and phase qubits), RF SQUIDs and many-junction superconducting quantum interferometers (flux qubits), just to name a few.
A crucial property of such systems is that at low temperatures they can be modeled as quantum-mechanical two-state systems displaying coherent quantum dynamical phenomena, i.e. quantum beating between two states \cite{FriedmanLukens,Mooij,IBM, Ustinov}, and, in the presence of externally applied radiation, microwave induced Rabi oscillations, Ramsey fringes etc. \cite{Martinis,Devoret,Ustinov2}. For single qubits these effects have been analyzed theoretically and observed experimentally.

As we turn to diverse systems containing  many interacting qubits quantum dynamics becomes more complex and interesting. First of all  due to a spread of parameters of individual qubits they perform quantum beating oscillations with different frequencies equal to (in the non-interacting case) $\omega_i=\Delta_i/\hbar$, where $\Delta_i$ is the energy level splitting of a single qubit. E.g. in Ref. \cite{Ust-TL} a system of seven flux qubits, i.e. three-junction superconducting quantum interferometers, has been studied to reveal a behavior corresponding to presence of seven different two-level systems.  Thus, the presence of unavoidable spread of parameters of qubits results in a non-synchronized quantum dynamics of non-interacting qubits. Similar results have been also obtained for a single Josephson junction containing a large amount of microscopic two-levels systems randomly distributed in its insulator interlayer \cite{Martinis-TLS,Ustinov-TLS}. Therefore, one could ask:
is it possible to observe \emph{collective quantum coherent phenomena} arising in the whole system?

In order to obtain such synchronized behavior in systems of many qubits an interaction between them has to be provided. It is well known that a strong long-range interaction between well-separated qubits can be induced by emission and absorption of virtual photons. This type of interaction was proposed in Refs. \cite{Wallr,Zagoskin,FU1,FU2,Fglstate} and realized in experiments with single qubits incorporated into a resonator \cite{Wallr2,Mooij2}. Moreover, measurements of frequency dependent transmission (reflection) coefficient of electromagnetic field propagating in the transmission line coupled to qubits provide a convenient method to observe coherent quantum phenomena in large systems of interacting qubits \cite{Ust-TL,Abdumalikov}.

In this Letter we show that in the presence of such interaction an array of $N$ qubits  displays \emph{two } collective coherent quantum oscillations. These quantum oscillations are characterized by two frequencies, $\omega_1=\bar{\Delta}/\hbar$ and $\omega_2=\tilde{\omega}_R$, where $\bar{\Delta}$ is the energy levels difference averaged over an ensemble of qubits, and $\tilde{\omega}_R$ is the resonator frequency renormalized by interaction.  Moreover, we obtain that the amplitude of these oscillations can be strongly enhanced in the resonant case as $\omega_1 \simeq \omega_2$.

In order to carry out the quantitative analysis of collective coherent quantum phenomena we consider a particular example of an array of $N$ RF SQUIDs inductively coupled to a resonant cavity. Each RF SQUID is characterized by a dynamic variable --- Josephson phase $\varphi_i(t)$. Potential relief for the Josephson phase $U(\varphi_i)$ can be tuned by externally applied magnetic field to have a double-well form. The resonator is characterized by two parameters $L_0$ and $C_0$, the inductance and capacitance per unit length, accordingly. The resonator frequencies are written as $\omega_R=ck_n$, where $c=1/\sqrt{L_0C_0}$ and $k_n=\pi n/\ell$, where $\ell$ is the size of the transmission line, $n=1,2...$. As the resonator has an extremely high quality factor only one wave vector will be important in the dynamics of coupled qubits and photons of resonator. Mutual inductance $M$ provides an interaction between RF SQUIDs and resonator. The schematic of such a system is presented in Fig. 1.

\begin{figure}
\includegraphics[width=1.5in,angle=-90]{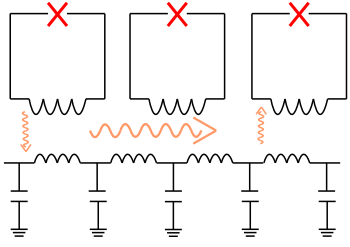}
\caption{The schematic of an array of RF SQUIDs incorporated into a resonator. An interaction through emission (absorption) of virtual photons is shown.
} \label{schematic}
\end{figure}

We start our quantitative analysis with the partition function $Z$ written as a path-integral over Josephson phases $\varphi_i(\tau)$, and the charge variable characterizing photon states in the resonator $Q(\tau)$, where $\tau$ is the imaginary time, i.e.
\begin{equation} \label{PartFunction}
Z=\int D[\varphi_i, Q] \exp \{-S[\varphi_i,Q]/\hbar \},
\end{equation}
where the action $S[\varphi_i,Q]$ is
$$
S[\varphi_i,Q]=\int_0^{\hbar/(k_BT)} d\tau [L_{qubits}+L_{res}+L_{int}]
$$
$$
L_{qubits}=E_J \sum_i \left\{
\frac{(\dot \varphi_i)^2}{2 \omega_p^2} -\frac{\alpha_i \phi_i^2}{2}+ \frac{\phi_i^4}{24}
\right\}
$$
$$
L_{res}=\frac{m}{2} [\dot Q^2 +\omega_R^2 Q^2], ~~m=L_0\ell/2
$$
\begin{equation} \label{Lagrangian}
L_{int}=iE_JQ\sum_i \eta_i \dot \varphi_i
\end{equation}
Here, $E_J$ and $\omega_p$ are the Josephson coupling energy and the plasma frequency, accordingly. The parameters $\eta_i$ and $\alpha_i$ have to be determined from the microscopic analysis. For our particular case of RF SQUIDs incorporated into a low-dissipation resonator the parameters have been obtained explicitly in Ref. \cite{MukhinFist}. Next, we trace out \cite{Feynman} the partition function over the charge variable $Q(\tau)$ and obtain the effective action $S_{eff}$ of N globally coupled two-level systems:
$$
S_{eff}=E_J
\sum_{i=1}^{N}\int_0^{\hbar/(k_BT)} d \tau \left[
    \frac{(\dot \varphi_i)^2}{2 \omega_p^2} -\frac{\alpha_i \varphi_i^2}{2}+ \frac{\varphi_i^4}{24} \right]+
$$
\begin{equation}
+\frac{E_J}{2} \sum_{i,j} \xi_i \xi_j\int_0^{\hbar/(k_BT)} d \tau \int_0^{\hbar/(k_BT)} d \tau'
    G_T(\tau-\tau')  \dot \varphi_i \dot \varphi_j,
\label{seff}
\end{equation}
where the kernel $G_T(\tau)$ is determined as
\begin{eqnarray}
G_T(\tau)=\frac{k_BT}{\hbar}\sum_n \frac{ e^{i\omega_n
\tau}}{\omega_n^2+\omega_R^2},~~~~~~~~~\nonumber \\
\omega_n=n(\pi k_BT)/\hbar,~ n=0, \pm 1,\pm 2....;
\label{kernel}
\end{eqnarray}
and $\xi_i=\eta_i \sqrt{E_J/m}$ are dimensionless coupling constants. Notice that a similar effective action has been used in Ref. \cite{Fglstate} in order to analyze macroscopic quantum tunneling in a globally coupled array of Josephson junctions.

\begin{figure}
\includegraphics[width=1.5in,angle=-90]{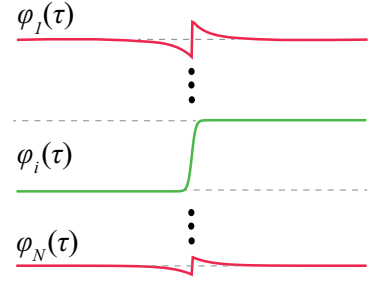}
\caption{The typical solution consisting of the single instanton "step" on $i$-th qubit
(solid green (thick) line) and tails on other qubits (solid red (thin) lines).
} \label{instanton}
\end{figure}

The partition function is determined by saddle point solutions that satisfy the equation:
\begin{equation}
\frac{\ddot \phi_i}{\omega^2_p}+\xi_i \sum_j \xi_j \int_0^{\beta \hbar}  G_T(\tau-\tau') \ddot \phi_j(\tau') d \tau' + \alpha_i \phi_i - \frac{\phi_i^3}{6} = 0.
\label{effeq}
\end{equation}
Let us consider solutions of the following form: an instanton (anti-instanton) solution $f(\tau)$ on $l$-th qubit and perturbative "tails" on other qubits. This type of saddle point solutions is shown in Fig. 2.
For the "tails" we linearize equations near the minimums of the potential $U(\varphi_i)$. In the absence of interaction the instanton (anti-instanton) solution is written as
\begin{equation}
f(\tau)=f_0(\tau) = \pm \sqrt{6 \alpha_l} \tanh \left[{\sqrt{\frac{\alpha_l}{2}} \omega_p (\tau-\tau_c)}\right],
\label{Instsol}
\end{equation}
where $\tau_c$ is the instanton "center" time. Using Eqs. (\ref{effeq}) and (\ref{Instsol}) we obtain the effective action of this solution as
\begin{equation}
S^l_{eff}= S^{l}_0+\frac{1}{2} \xi_l^2 \int_0^{\beta \hbar} d \tau \int_0^{\beta \hbar} d \tau'
    G_1(\tau-\tau')  \dot f_0(\tau) \dot f_0(\tau'),
\label{NewSeff}
\end{equation}
where $S^{l}_0$ is the action calculated for instanton solution $f_0(\tau)$ in the absence of interaction between qubits, and the Fourier transform of the kernel $G_1(\omega_n)=G_T(\omega_n)/[1+\kappa(\omega_n)]$, where $\kappa(\omega_n) = \sum_{j \neq l} \frac{\eta_j^2 \omega^2 G_T(\omega_n)}{\omega_n^2/\omega_p^2+2 \alpha_j}$.
Now considering multi-instanton solutions with the help of \emph{non-interacting instanton (anti-instanton)} approximation \cite{Leggett}, which is valid for rather weak interaction strength between qubits,  one can calculate the partition function $Z$ in a similar fashion to \cite{Chakr}.
It is then written as $Z=\prod_{i=1}^N Z_i$, $Z_i=2\cosh [\Delta_i/(k_BT)]$ and $\Delta_i~\simeq~\frac{\hbar}{\omega_p\sqrt{\alpha_i}}\exp(-S^i_{eff}/\hbar)$. Calculating integrals over $\tau$ in Eqs. (\ref{seff}) and (\ref{NewSeff}) we obtain
\begin{equation}
 S^{i}_{eff}=E_J 4 \sqrt{2} \frac{\alpha_i^{3/2}}{\omega_p}+S^i_{int}~.
\label{Seff0}
\end{equation}
The coupling between qubits results in an enhancement of effective action $S^{i}_{eff}$ , and therefore, a decrease of average level splitting $\Delta_i$. Moreover, the dispersion of qubit level splittings will be enhanced. The explicit value of $S_{int}$ is determined by the parameter
\begin{equation}
\beta ~\simeq~ [2+ (N-1) \langle \eta_j^2/\alpha_j \rangle]
\label{betaeq}
\end{equation}
and the ratio of two frequencies: $\sqrt{\alpha_i}\omega_p$, i.e. the frequency of small oscillations on the bottom of potential well, and $\omega_R$.
Here, the $\langle ...\rangle$ determines the averaging over a spread of qubits parameters $\xi_i$, $\alpha_i$ and $\Delta_i$.
Explicit calculating integrals in \ref{NewSeff} allows one to  obtain
\begin{equation}
S^i_{int}=\xi_i^2\alpha_i \frac{2\sqrt{2}E_J }{\omega_R}
\begin{cases}
\frac{3}{ \sqrt{\beta}}
& \text{if }
\beta \left(\frac{\sqrt{\alpha_i}\omega_p}{\omega_R} \right)^2
\gg 1
\\
\frac{\sqrt{\alpha_i}\omega_p}{\omega_R}
& \text{if }
\beta \left(\frac{\sqrt{\alpha_i}\omega_p}{\omega_R} \right)^2
\ll 1
\end{cases}
\end{equation}

Next, in order to analyze the quantum dynamics of an array of interacting qubits we obtain the time-dependent correlation function of a single qubit, i.e. $C_i(t)=<\varphi_i(t) \varphi_i(0) > $. In the non-interacting instanton (anti-instanton) approximation we can write $\varphi_i(\tau)$ as a sum:
\begin{equation}
\varphi_i(\tau) = f_i(\tau) + \sum_{j \neq i} \widetilde \varphi_i^j(\tau),
\label{phase1}
\end{equation}
where $f_i(\tau)$ consists of instanton (anti-instanton) "kinks" and $\widetilde \phi_i^j(\tau)$ correspond to "tails" from instantons (anti-instantons) on $j$-th qubit. The typical solution $\varphi_i(\tau)$ for a single instanton  and many instantons (anti-instantons) are shown in Figs. 2 and 3.

The correlation function $C_i(\tau)$ is written as
\begin{equation}
C_i(\tau)=<f_i(\tau)f_i(0) > +\sum_{j \neq i} < \widetilde \varphi_i^j(\tau)\widetilde \varphi_i^j(0)>.
\label{Korrfunction}
\end{equation}
Following the Ref. \cite{Chakr} the first term in the right-hand part of Eq. (\ref{Korrfunction}) is obtained as
$$
C^{0}_i(\tau)=<f_i(\tau)f_i(0) >=\frac{2}{Z} \sum_{n=0}^{\infty} \left(\frac{\Delta_i}{\hbar}\right)^{2n}
$$
\begin{equation}
\int_0^{\hbar/(k_BT)} d t_{2n} \int_0^{t_{2n}} d t_{2n-1} ... \int_0^{t_2} d t_1 \prod_{i=1}^{2n} \varphi_0^2 \text{sgn}(t_i - \tau),
\label{KF-NO-Int}
\end{equation}
where $\pm \varphi_0=\pm \sqrt{6\alpha_i}$ are the minima of the double-well potential $U(\varphi_i)$. Calculating the integrals over $t_n$ (the instanton "center" times) we obtain
\begin{equation}
C^{0}_i(\tau)=\varphi_0^2 \frac{\cosh([\frac{\hbar}{k_BT} - 2 \tau]\Delta_i)}{\cosh [\Delta_i/(k_BT)]}
\label{KF-NO-Int-2}
\end{equation}
Carrying out the analytical continuation to the real time we obtain in the low-temperature limit, i.e.  $k_B T<<\Delta_i$, the correlation function of non-interacting qubits as
\begin{equation}
C^{0}_i(t)=\varphi_0^2 e^{-2i\Delta_i t/\hbar}
\label{KF-Noint}
\end{equation}
This result indicates presence of quantum beating oscillations with $N$ different frequencies, $\omega_i=\Delta_i/\hbar$, in the system.

However, there is another contribution to the correlation function of $i$-th qubit stemming from the tails of instantons (anti-instantons) occurring on other qubits. Such a contribution shown in Figs. 2 (a single instanton solution) and 3 (many instanton (anti-instanton) solution), is written as
\begin{equation}
\widetilde \varphi_i^l(\tau) = \frac{k_B T}{2 \pi \hbar}\sum_n \int d\tau_1 G(\omega_n) e^{i\omega_n(\tau-\tau_1)}f_l(\tau_1),
\label{Insttail}
\end{equation}
where
\begin{equation}
G(\omega)= - \frac{\xi_i \xi_l \omega^2}{(2\alpha_i+(\omega/\omega_p)^2)[\omega^2+\omega_R^2+(N-1)\langle \xi_j^2/(2 \alpha_j) \rangle \omega^2]}
\label{Grfunction}
\end{equation}
Substituting (\ref{Insttail}) in (\ref{Korrfunction}) and taking into account  that $C_l^0 (\tau)= <f_l(\tau)f_l(0) >$ we obtain
\begin{equation}
<\widetilde \phi_i^l(\tau) \widetilde \phi_i^l(0) > = \varphi_0^2\int d\tau_1 \sum_n G(\omega_n) G(-\omega_n)e^{i\omega_n(\tau-\tau_1)}C_l^0 (\tau_1)~.
\label{Korrfunction-Inter}
\end{equation}
The quantum-mechanical dynamics is determined by the renormalized frequency of the resonator $\widetilde{\omega}_R=\omega_R\sqrt{2/\beta }$. In the limit of $\frac{\sqrt{\alpha_i}\omega_p}{\widetilde{\omega}_R} \gg 1$ the kernel $G(\omega)$ is simplified as $G(\omega)=- \frac{\xi_i \xi_l}{\beta \alpha_i}\frac{ \omega^2}{[\omega^2+\widetilde\omega_R^2]}$.

\begin{figure}
\includegraphics[width=1.7in,angle=-90]{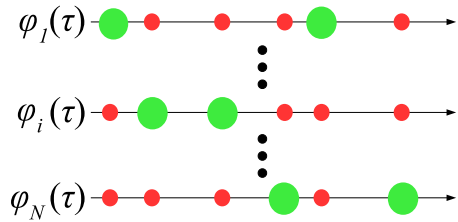}
\caption{A solution consisting of large amount of instantons (anti-instantons) (large (green) circles) and corresponding tails (small (red) circles) is shown.
} \label{instanton-large}
\end{figure}

Carrying out the analytical continuation to the real time \cite{Ingold} we obtain the time-dependent correlation function in the following form:
$$
C_i(t)=C_i^0(t)+C_1^{col}(t)+C_2^{col}(t),
$$
where the time-dependent correlation functions $C_{1,2}^{col}(t)$ are expressed as
$$
C_1^{col}(t)=(N-1)\varphi_0^2\left(\frac{\xi_i^2}{\alpha_i \beta}\right)^2
$$
\begin{equation}
\left \langle \frac{16\xi_l^2 (\Delta_l/\hbar)^4}{(\widetilde \omega_R^2-4(\Delta_l/\hbar)^2)^2+4 \gamma^2(\Delta_l/\hbar^2)} e^{-\frac{2i\Delta_l}{\hbar}(t)}
    \right \rangle,
\label{Collectiveexcitation-1}
\end{equation}
and
$$
C_2^{col}(t)=(N-1)\varphi_0^2\left(\frac{\xi_i^2}{\alpha_i \beta}\right)^2 e^{-[2i\widetilde{\omega}_R+\gamma/2]t}
$$
\begin{equation}
\left \langle \frac{-i 16\xi_l^2 (\Delta_l/\hbar)\widetilde{\omega}_R^2/\gamma}{4(\Delta_l/\hbar)^2-\widetilde \omega_R^2+i\gamma\widetilde \omega_R/2 }
    \right \rangle,
\label{Collectiveexcitation-2}
\end{equation}
where $\gamma$ is a phenomenological parameter describing dissipation in the system. This parameter allows one to keep finite the resonant term in the correlation functions $C_{1,2}^{col}(t)$  as $\widetilde{\omega}_R~\simeq~\Delta_l$.  The correlation functions $C_{1,2}^{col}(t)$ determine two \emph{collective quantum-mechanical oscillations } with two frequencies, namely, the energy level splitting averaged over an ensemble of qubits, $\omega_1=\bar{\Delta}/\hbar$ and self-frequency of the resonator renormalized by interaction $\omega_2=\widetilde{\omega}_R$. These collective oscillations are excited by coherent quantum beatings in a system of globally coupled qubits.  Moreover, oscillations with frequency $\omega_1$ decay in time due to the dissipation and to a spread of qubits parameters. The second type of oscillations with the frequency $\omega_2$ decays in time due to the dissipation effects only. The amplitudes of these oscillations enhance strongly in the resonant case as $\widetilde{\omega}_R~\simeq~\Delta_l$. Such an enhancement can also lead to a suppression of the double-well potential barrier for the Josephson phase, and, therefore, to an increase of level splittings $\Delta_i$. This effect is similar to a well-known microwave induced enhancement of macroscopic quantum tunneling in Josephson junctions \cite{WFU-IMQT}.

In conclusion, we have shown that an array of strongly coupled qubits can display coherent collective quantum oscillations.
We consider a particular example of an array of superconducting qubits (RF SQUIDs) incorporated into a resonator. In such a system a long-range interaction (a global coupling) can be provided by emission (absorption)  of virtual photons in the resonator. In the presence of such interaction we obtain a decrease of average qubit levels splitting.  The dispersion of qubit level splitting  is enhanced. However, by analyzing quantum-mechanical correlation functions we obtain that beyond quantum beating oscillations with different frequencies, $\omega_i=\Delta_i/\hbar$, there are two collective quantum-mechanical oscillations with two frequencies, $\omega_1$ and $\omega_2$. These collective oscillations appear in the presence of a long-range coupling between qubits, and they are induced by coherent quantum beatings occurring in whole system. In order to observe these collective oscillations the temperature has to be low, i.e. $k_BT<< \hbar \omega_{1,2}$, the dissipative effects small, and spread of parameters $\Delta_i$ not large. Such coherent collective quantum oscillations can be observed either in artificially prepared arrays of qubits incorporated in the low-dissipation resonator or in single Josephson junctions containing a large amount of microscopic two-level systems. The observation of these collective quantum-mechanical modes  will provide an evidence of synchronized quantum dynamics in a system of strongly interacting qubits.

We acknowledge partial support of this work by the Russian Ministry of Science and Education grant No. 14A18.21.1936. P. A. V. acknowledges the financial support of Russian Quantum Center (RQC) and the hospitality of the Ruhr-University Bochum where this work has been made.

{}

\end{document}